# Simulation and analysis of turbulent flame and its effect on the wall of aero-engine combustor

**A.Mokhtari[1], A.Abdallah-elhirtsi[2], F.Larbi[3], R.Renane[4*], R.Allouche[5]**

r.renane@gmail.com*

[1,2,3,4,5] Laboratory of aeronautical sciences, Institute of aeronautics & space studies,
*University of Blida 1, BP 270 Blida 09000, Algeria.*

## ABSTRACT

**Abstract:**

The main objective of this study is to simulate the behavior of the reactive flow of the turbulent flame in aeronautical combustion chamber of the ALLISON-T56 turboprop, and contribute to the analysis of flame structure and determine for given pressure and temperature of fresh gas the behavior of the thermodynamic parameters of combustion. The numerical approach is based on the resolution of basic equations of turbulent combustion using Ansys-Fluent code where the turbulence model K-ε is chosen, the geometry of the combustion chamber is made using Ansys-workbench software. Thereafter, we simulate the transient temperature field through the wall of a tubular combustion chamber, and the characterization of the thermal expansion, the thermoelastic stresses and strains with the physical properties of refractory materials. The obtained results are then compared with the results of the scientific literature.

Corresponding author: r.renane@gmail.com

## NOMENCLATURE

**Symbols :**
k    turbulente energie
T    temperature, K
P    pressure, $Nm^{-2}$

**Greek Letters:**
$\rho$ density, $kgm^{-3}$
$\nu$ eddy viscosity

P the production rate of turbulent kinetic energy
$\varepsilon$ dissipation
$U^+$ speed profil
$\tau_{max}$ Maximum shear stress
$\sigma_e$ Equivalent stress
$\sigma_p$ Main stress
$Y^+$ distance of the first cell adjacent to the wall

## 1. INTRODUCTION

Combustion is one of the means of energy conversion, characterized by a highly exothermic irreversible reaction between an oxidant and a fuel. The study of this phenomenon has a considerable interest in the aviation sector [1]. The main preoccupation of manufacturers and researchers is to master the behavior of various thermodynamic parameters for efficient and ecological combustion [3]. Our work is a contribution to the analysis and simulation of the structure of the turbulent diffusion flame in annular combustion chamber of turboprop ALLISON-T56. The fuel used in our study is kerosene, which has a high calorific value of 43.15 MJ.kg-1. In this work we are interested first to the geometry of the combustion chamber studied; thereafter, we present the discretization of the computational domain by the mesh generation using the Gambit software (Figure 2). The equations governing the gaseous reactants flows are recalled in the previous section where a brief reminder of the k-e turbulence model used is presented. The simulation results are presented and discussed in the last section, and a comparison is made with the scientific literature. On the other hand, the temperature reached in the aeronautical combustion chambers exceeds the limit characteristics of thermal resistance of current materials; it's for this reason that the temperature at the end of combustion must be controlled. The thrust of the engine is directly related to the temperature of gas emissions [5]. We conceive the importance of having materials resistant to high temperatures, which have good mechanical strength and corrosion resistance, that they can supporting overheating without the risk of weakening before high stresses (creep), and finally, to be elaborate without need for heat treatment.

## 2. MATHEMATICAL MODEL

For this study, calculations are executed by using the Ansys software, where several models of turbulence are available in this code, the models with one and two transport equations use partial derivative equations to connect the fluctuations of flow to the average sizes of variables [6]. We limit as an example to present thereafter the K-ε model. The K-w model and SST- Model are respectively detailed in references [5-7]. In the second section, the Ansys-Fluent code is used again to simulate the interaction of the turbulent flame on the structure and the thermoelastic behavior of the wall of the combustion chamber; a brief reminder is given of the thermoelastic theory developed by W.D.KINGERY.

2.1    k-ε model

The k-ε model [2,7] is a model with two transport equations to evaluate the vortex viscosity. It solves two partial differential equations for the turbulent kinetic energy k and its dissipation ε, these equations are:

$$\frac{\partial}{\partial t}\left(\overline{\rho}k\right)+\frac{\partial}{\partial x_j}\left(\overline{\rho}u_j k\right)=\overline{\rho}P-\overline{\rho}\varepsilon+\frac{\partial}{\partial x_j}\left[\left(\overline{\mu}+\frac{\overline{\mu}_t}{p_{rk}}\right)\frac{\partial k}{\partial x_j}\right] \quad (1)$$

$$\frac{\partial}{\partial t}\left(\overline{\rho}\varepsilon\right)+\frac{\partial}{\partial x_j}\left(\overline{\rho}u_j \varepsilon\right)=C_{\varepsilon 1}\frac{\overline{\rho}P\varepsilon}{k}-C_{\varepsilon 2}\frac{\overline{\rho}\varepsilon^2}{k}+\frac{\partial}{\partial x_j}\left[\left(\overline{\mu}+\frac{\overline{\mu}_t}{P_{r\varepsilon}}\right)\frac{\partial \varepsilon}{\partial x_j}\right] \quad (2)$$

Where P is the production rate of turbulent kinetic energy given by the equation:

$$P=\overline{\upsilon}_t\left(\frac{\partial u_i}{\partial x_j}+\frac{\partial u_j}{\partial x_i}-\frac{2}{3}\frac{\partial u_m}{\partial x_m}\delta_{ij}\right)\frac{\partial u_i}{\partial x_j}-\frac{2}{3}k\frac{\partial u_m}{\partial x_m} \quad (3)$$

With the constants are given by [7]:
$$C_\mu = 0.09, C_{\varepsilon 1} = 1.44$$
$$C_{\varepsilon 2} = 1.92, P_{rk} = 1.0 \; et \; P_{r\varepsilon} = 1.3$$

Viscous effects are greater than the turbulent effects in the vicinity of the wall. A wall law (Wall) function is, therefore, applied in this region and the turbulence model solves the field in the rest of the field of flow. In the law of the wall, the scale of the speed is taken as q=k$^{0.5}$, and scale length is modeled by $l = \dfrac{C_\mu^{3/4} k^{3/2}}{\varepsilon}$.

k, and ε are connected by semi-empirical expressions for the friction velocity as follows:

$$k = \frac{U_\tau^2}{\sqrt{C_\mu}} \quad and \quad \varepsilon = \frac{C_\mu^{3/4} k^{3/2}}{\kappa y} \tag{4}$$

Eddy viscosity is expressed by:
$$\upsilon_t = \frac{C_\mu k^2}{\varepsilon} \tag{5}$$

In the standard approach of the law of the wall), the speed profile is estimated from the first wall to the mesh by the following equations [7]:

$$U^+ = y^+ \quad pour \quad y^+ < 11.5$$
$$U^+ = \frac{1}{\kappa} \ln(Ey^+) \quad pour \quad y^+ > 11.5 \tag{6}$$

Where the constants k,E are determined experimentally and are worth: 0.4 and 9.0 respectively. The concept wall law is valid in the case where the value of the center distance of the first cell adjacent to the wall is such that $y^+ > 30$.

2.2 Thermoelastic behavior

Thermal stresses, sometimes referred to as "thermal stresses", come from the fact that a material subjected to a change in temperature is constrained in such a way that it cannot deform freely. In this case, the thermal deformation is compensated by elastic deformation.
   It is therefore the elastic deformation (which corresponds to a displacement of the atoms with respect to their equilibrium position, which has changed with temperature) which is at the origin of the stress, it is only indirectly that the temperature change induces a stress [4].

2.2.1 Elastic limit criteria

   In a one-dimensional tensile test, the elastic limit is defined as the stress for which the first plastic deformations appear. Below this limit, all deformations generated during the loading of the specimen can be recovered. This definition of the elastic domain for an axial plain test must be generalized in the case of complex loading. This three-dimensional generalization is called the plasticity criterion. It allows to define. In the space of the stresses, the region for which the material will have an elastic behavior. The definition of the two most widely used isotropic criteria for metals, the Tresca and Von Mises criteria is given below.

2.2.2 Criterion of Tresca

The maximum shear stresses are given by the expressions:

$$\begin{cases} \tau_{max12} = \frac{1}{2}(\sigma_{p1} - \sigma_{p2}) \\ \tau_{max13} = \frac{1}{2}(\sigma_{p1} - \sigma_{p3}) \\ \tau_{max23} = \frac{1}{2}(\sigma_{p2} - \sigma_{p3}) \end{cases} \quad (7)$$

Where: $\tau_{max}$ Maximum shear stress and $\sigma_p$ Main stress

The material must meet the following strength requirements:

$$\sigma_{max} \leq [\sigma]_{adm} \quad \text{and } \tau_{max} \leq \frac{[\sigma]_{adm}}{2} \quad (8)$$

Where: $\sigma_{adm}$ allowable stress

2.2.3 Criterion of Von Mises

There exists another criterion to check the resistance condition, it is the one given by Von-Mises, it defines the equivalent stress by:

$$\sigma_e = \sqrt{\frac{1}{2}[(\sigma_{p1} - \sigma_{p2})^2 + (\sigma_{p1} - \sigma_{p3})^2 + (\sigma_{p3} - \sigma_{p2})^2]} \quad (9)$$

The material must meet the following strength requirement: $\sigma_e \leq [\sigma]_{adm}$

$\sigma_e$ Equivalent stress

## 3. RESULTS AND DISCUSSION

Among the obtained results, we shows on the fig.2 the mesh of combustion chamber realized by Gambit software, and we observe on the fig.3 the isotherm areas in the combustion chamber. We notice a flame is formed, and one can easily see asymmetry that is quite good. The temperature is maximum at the center which is quite logical and the luminescence temperature is between 1200and 1500 °C. According to figure (4), we see a great similarity not only as regards the shape and symmetry of the flame which is centralized by the air holes whose temperature distribution peaked at about the 2300K, 850 K on the wall and 1300 K at the exit of the chamber. We also note that the distribution of temperature in the combustion chamber reached a maximum of 2306 K and a progressive decrease is observed in the direction of flow to the exit of the chamber but also the center to the wall cooled by a protective film entering through different cooling holes. The flame can be divided into three area according to the temperature variation. 1.reaction area, 2. Secondary area and 3. Dilution area.

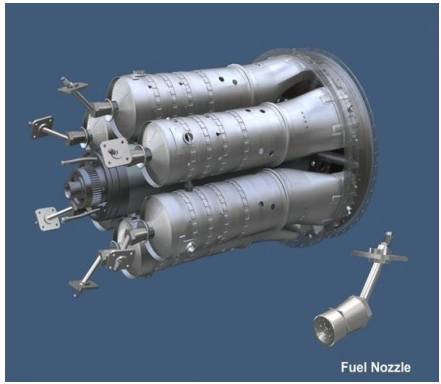
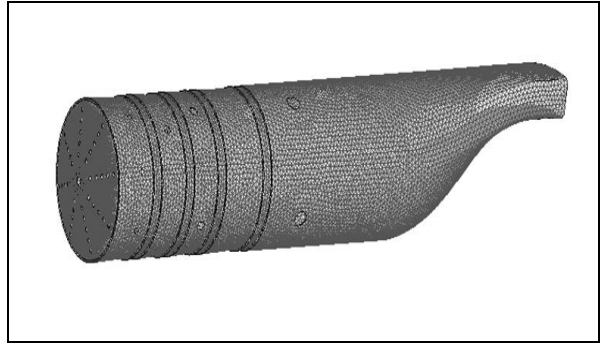

Fig. 1. Real combustion chamber [8]     Fig. 2. Mesh of combustion chamber

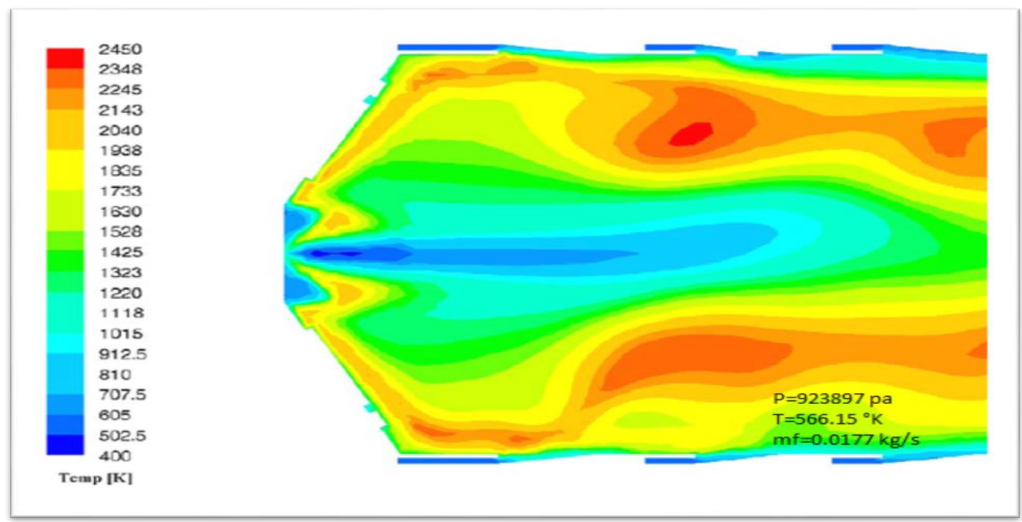

Fig.3 Simulation of turbulent flame in the combustion

FIG. 4 illustrates the influence of the preheating temperature of the air / fuel mixture on the combustion temperature along the chamber. Concerning the maximum temperature, it is noted that the increase in the inlet temperature raises the maximum temperature, thus for a preheating temperature of 450 K; The Max temperature is of the order of 2050k, whereas for an input temperature of 566k, the Tmax = 2260k. As well as for that of the outlet by increasing the inlet temperature, the outlet temperature varies from 1230 K to 1360 K. This allowed us to conclude that the preheating of the air / fuel mixture improves combustion efficiency [4,8].

FIG. 5 shows the variation of the turbine inlet temperature (TIT) as a function of the fuel flow rate. It can be seen in the figure that the TIT increases linearly between 1150k and 1400k with the increase in the fuel flow rate. We note that our results obtained by the simulation correlate with the results that we have obtained on the turboprop test bench of the (Air Algerie) company.

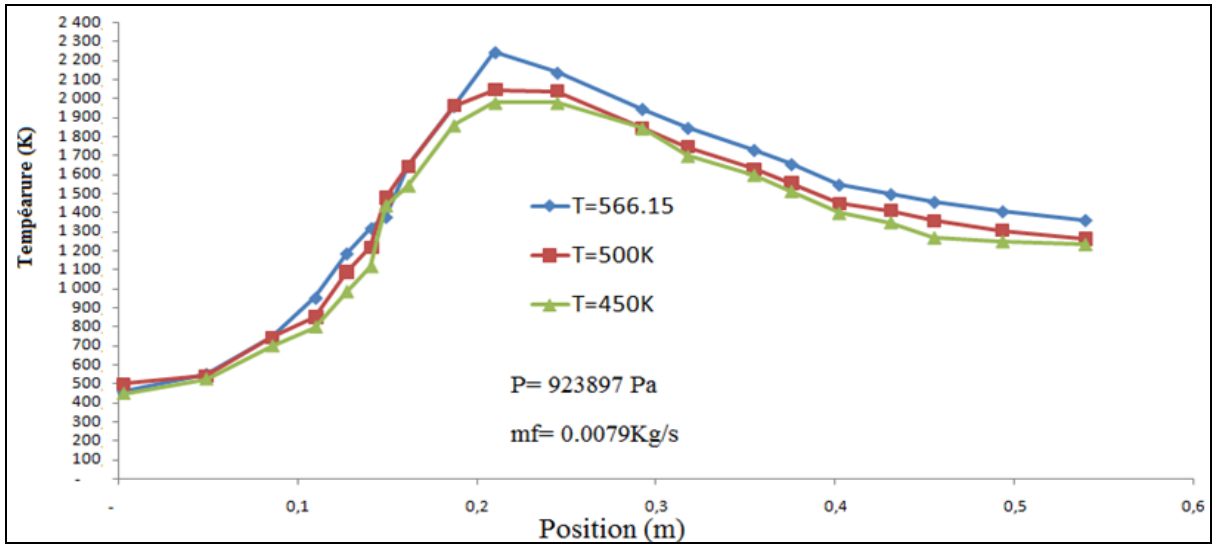

Fig.4 Distribution of the temperature in the combustion chamber as a function of the preheating temperature.

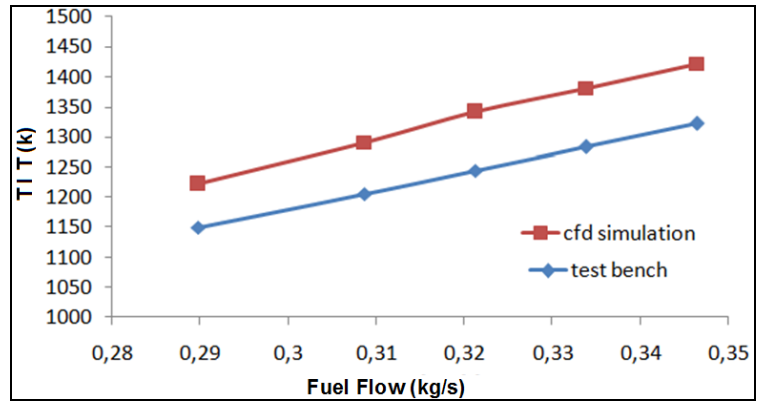

Fig.5 Change in TIT (turbine inlet temperature) as a function of fuel flow

Fig.6 and Fig.7 indicate respectively, the variation of the equivalent stress with Von-Mises criterion and the variation of the maximum shear stress with Tresca criterion for the same Titanium alloy, we note that both the equivalent stress and the maximum shear stress follow the same behavior as the temperature through the combustion chamber (FIG. 5), so they start to increase in the primary zone of the chamber until they reach a maximum and then they gradually decrease in the secondary zone of the chamber where the temperature stabilizes at the TIT.

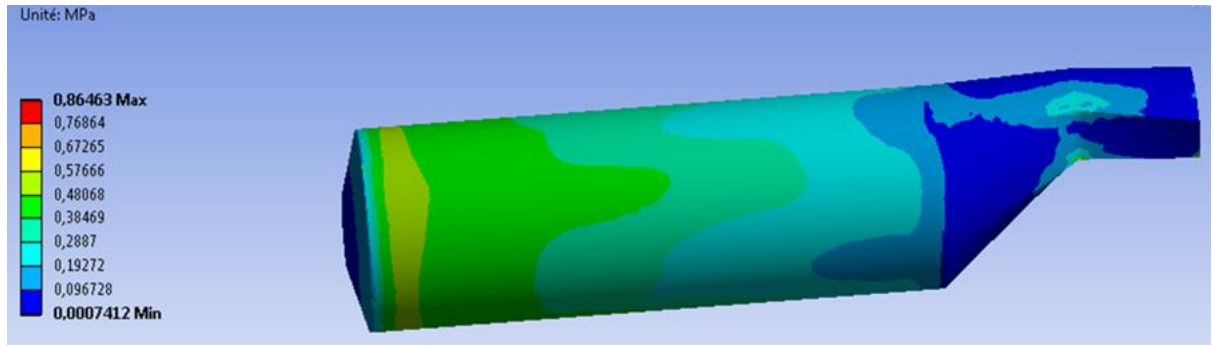

Fig.6 Variation of the equivalent stress with Von-Mises criterion for a Titanium alloy

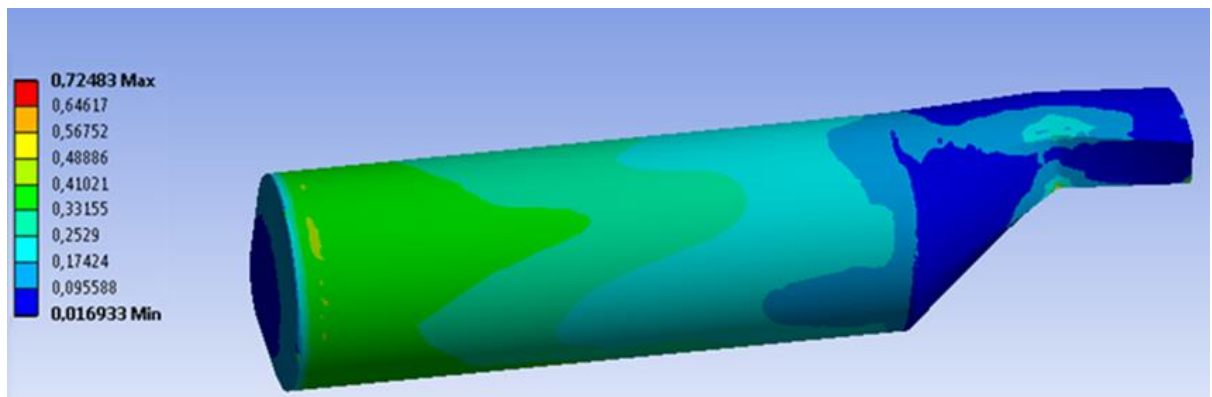

Fig.7 Variation of the maximum shear stress with Tresca criterion for a Titanium alloy

## 4. CONCLUSIONS

In this study devoted to the numerical simulation of the turbulent diffusion flame in the combustion chamber of the engine Allison T56, it has been realized the mesh of the rather complex geometry of the combustion chamber using both the Gambit software and Ansys –Fluent, a mesh sensitivity study is made for the choice of optimum mesh. Furthermore this study allowed us to see the influence of the injection gas temperature and the initial pressure on the flame structure. We can easily see that the symmetry of the flame formed is quite good, the temperature is maximum at the center which is quite logical and the luminescence temperature is between 1200 and 1400 ° C. In addition, this study allowed us to conclude that the preheating of the air / fuel mixture and the incrising of initial pressure improves combustion efficiency [8]. Moreover the turbine inlet temperature TIT increases with the increase in fuel flow. Furthermore, the variation of the equivalent stress with Von-Mises criterion and the variation of the maximum shear stress with Tresca criterion for the same Titanium alloy follow have the same behavior as the temperature through the combustion chamber.